\documentclass[aps,prl,twocolumn,superscriptaddress,showpacs,preprintnumbers,amsmath,amssymb,nofootinbib]{revtex4-2}

\usepackage{graphicx} % Include figure files
\usepackage{dcolumn}  % Align table columns on decimal point
\usepackage{soul} 
\usepackage{dblfloatfix}
\usepackage{multirow}
\usepackage{amsmath}
\DeclareUnicodeCharacter{2212}{\textendash}
\graphicspath{{ps}}

\usepackage [autostyle, english = american]{csquotes}
\usepackage{ifthen} % for conditional statements
\usepackage{orcidlink} % for ORCIDs in author list

\newboolean{articletitles}
\setboolean{articletitles}{true} % False removes titles in references

\MakeOuterQuote{"}
\usepackage{hyperref}
\hypersetup{
    colorlinks,
    citecolor=blue,
    filecolor=blue,
    linkcolor=blue,
    urlcolor= blue,
    }
\ExplSyntaxOn
\NewDocumentCommand{\longdash}{ O{2} }
 {
  --\prg_replicate:nn { #1 - 1 } { \negthinspace -- }
 }
\ExplSyntaxOff

\begin{document}

\title{Search for the decay $B^{0}\to\gamma\gamma$ using Belle and Belle II data}

%%% Paper:    B0 to gamma gamma
%%% Journal:  Physical Review D Letter
%%% Contacts: S.K. Maurya, D. Kalita, B. Bhuyan
%%% ====================================================================
%%% Use \input{pub051-orcid} to insert this material into your latex file.
  \author{I.~Adachi\,\orcidlink{0000-0003-2287-0173}} % 2590
% \author{K.~Adamczyk\,\orcidlink{0000-0001-6208-0876}} % 2239
  \author{L.~Aggarwal\,\orcidlink{0000-0002-0909-7537}} % 10083
% \author{P.~Ahlburg\,\orcidlink{0000-0002-9832-7604}} % 2367
% \author{H.~Ahmed\,\orcidlink{0000-0003-3976-7498}} % 11323
% \author{J.~K.~Ahn\,\orcidlink{0000-0002-5795-2243}} % 7423
  \author{H.~Aihara\,\orcidlink{0000-0002-1907-5964}} % 2223
  \author{N.~Akopov\,\orcidlink{0000-0002-4425-2096}} % 9443
  \author{A.~Aloisio\,\orcidlink{0000-0002-3883-6693}} % 2194
  \author{S.~Al~Said\,\orcidlink{0000-0002-4895-3869}} % 6823
  \author{N.~Althubiti\,\orcidlink{0000-0003-1513-0409}} % 21524
% \author{L.~Andricek\,\orcidlink{0000-0003-1755-4475}} % 2098
% \author{M.~Angelsmark\,\orcidlink{0000-0003-4745-1020}} % 13963
  \author{N.~Anh~Ky\,\orcidlink{0000-0003-0471-197X}} % 2218
  \author{D.~M.~Asner\,\orcidlink{0000-0002-1586-5790}} % 4684
  \author{H.~Atmacan\,\orcidlink{0000-0003-2435-501X}} % 2538
% \author{V.~Aulchenko\,\orcidlink{0000-0002-5394-4406}} % 8183
% ...
% 13403
\collaboration{The Belle and Belle II Collaborations}

\noaffiliation

\begin{abstract}
We report the result of a search for the rare decay $B^{0} \to \gamma \gamma$ using a combined dataset of \mbox{$753\times10^{6}$} $B\bar{B}$ pairs collected by the Belle experiment and $387\times10^{6}$ $B\bar{B}$ pairs collected by 
the \mbox{Belle II} experiment from decays of the $\rm \Upsilon(4S)$ resonance produced in $e^{+}e^{-}$ collisions. A simultaneous fit to the Belle and \mbox{Belle II} data sets yields $11.0^{+6.5}_{-5.5}$ signal events, corresponding to a 2.5$\sigma$ significance. We determine the branching fraction $\mathcal{B}(B^{0} \to \gamma\gamma) = (3.7^{+2.2}_{-1.8}(\rm stat)\pm0.5(\rm syst))\times10^{-8}$ and set 
a 90\% credibility level upper limit of $\mathcal{B}(B^{0} \to \gamma\gamma) < 6.4\times10^{-8}$.
\end{abstract}

% \pacs{XX.YY.ZZ, AA.BB.CC}

\maketitle
\tighten

{\renewcommand{\thefootnote}{\fnsymbol{footnote}}}
\setcounter{footnote}{0}

\begin{figure}[htb]
	\includegraphics[width=0.5\textwidth]{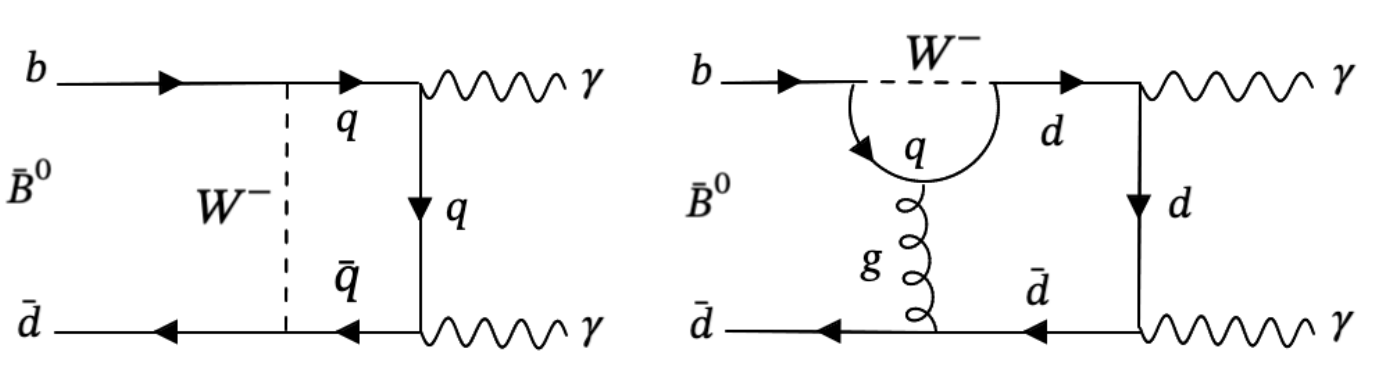}
	\caption{Box (left) and penguin (right) amplitudes contributing to $\overline{B}^{0}\to\gamma\gamma$ at leading order in the SM. The symbol $q$ represents a $u$, $c$, or $t$ quark.}
	
	\label{loop}
\end{figure}

In the standard model (SM), there is no tree-level interaction between the $b$ quark and the $d$ quark; therefore, the \mbox{$B^0 \rightarrow \gamma \gamma$} proceeds through a flavor changing neutral current transition involving electroweak loop amplitudes where a quark emits and reabsorbs a $W^-$ gauge boson. The dominant amplitudes are illustrated in Fig. \ref{loop}. 
% Possible Feynman diagrams contributing to this channel are shown in Fig \ref{loop}.
% This process is illustrated in Fig. \ref{loop}, where the $b$ quark undergoes a $b \rightarrow (u,c,t) \rightarrow d$ transition. 

The charge-conjugation and parity (CP) averaged branching fraction $\mathcal{B}(B^{0} \to \gamma\gamma)$ is predicted in the SM to be $(1.4^{+1.4}_{-0.8})\times10^{-8}$, including next-to-leading logarithmic and next-to-leading power corrections \cite{shen}. Although the long-distance penguin contribution is expected to be negligible, it can noticeably impact the CP-violating observables \cite{quin}.
The decay $B^{0} \to \gamma\gamma$ is sensitive to physics beyond the SM since contributions of non-SM particles in the loop could enhance the branching fraction by an order of magnitude to a few orders of magnitude above the SM expectation, depending on specific supersymmetric parameter values \cite{newphy,newphy2}. A measurement of $B^{0} \to \gamma\gamma$ thus offers an attractive opportunity to test theories beyond the SM \cite{aranda, newphy,newphy2}. 

The most stringent upper limit (UL) on the branching fraction is $\mathcal{B}(B^{0} \to \gamma\gamma) < 3.2 \times 10^{−7}$ at the 90\% confidence level (CL), set by the \textit{B\resizebox{!}{0.6em}{A}B\resizebox{!}{0.6em}{AR}} experiment \cite{del2011search} using an $e^{+}e^{-}$ dataset recorded at the $\rm \Upsilon(4S)$ resonance with an integrated luminosity of 426 $\text{fb}^{−1}$. The Belle experiment obtained the upper limit \mbox{$\mathcal{B}(B^{0} \to \gamma\gamma) < 6.2 \times 10^{-7}$} at the 90\% CL with a 104 fb$^{-1}$ $e^{+}e^{-}$ dataset from the $\rm \Upsilon(4S)$ \cite{belleul}.

Here, we report a search for the decay $B^0 \rightarrow \gamma \gamma$ using a combined $e^{+}e^{-}$ dataset from the Belle and \mbox{Belle II} experiments collected at the $\rm \Upsilon(4S)$ resonance energy. For Belle, we use a dataset corresponding to 694 $\text{fb}^{−1}$ containing $(753 \pm 10) \times 10^6$ $B\overline{B}$ pairs, while for \mbox{Belle II} we use 362 $\text{fb}^{−1}$ collected between 2019 and 2022, corresponding to $(387 \pm 6) \times 10^6$ $B\overline{B}$ pairs. The number of $B\overline{B}$ events from Belle used in this analysis is slightly smaller than that of the entire Belle dataset \mbox{(772 $\pm$ 11) $\times$ 10$^{6}$}, as we only use the data containing calorimeter timing information. The analysis does not distinguish between $B^{0}$ and $\overline{B}^{0}$, and throughout this article, charge conjugation is implied for all decays.

% {\it SVD1:}
% {The Belle detector is a large-solid-angle magnetic
% {spectrometer that
% {consists of a three-layer silicon vertex detector (SVD),
% {a 50-layer central drift chamber (CDC), an array of
% {aerogel threshold Cherenkov counters (ACC), % <- \v{C}erenkov 2007.08
% {a barrel-like arrangement of time-of-flight
% {scintillation counters (TOF), and an electromagnetic calorimeter
% {comprised of CsI(Tl) crystals (ECL) located inside 
% {a super-conducting solenoid coil that provides a 1.5~T
% {magnetic field.  An iron flux-return located outside of
% {the coil is instrumented to detect $K_L^0$ mesons and to identify
% {muons (KLM).  The detector
% {is described in detail elsewhere~\cite{Belle}.

% {\it SVD2+SVD1:}
%
%%%%%%%%%%%%%%%%%%%%%%%%%%%%%%%%%%%%%%%%%%%%%%%%%%%%%%%%%%%%%%%%%%5
%
%  NOTE: The Publication Council affirms (14/6/2021) that
%  this text is in the PRESENT TENSE.
%
%%%%%%%%%%%%%%%%%%%%%%%%%%%%%%%%%%%%%%%%%%%%%%%%%%%%%%%%%%%%%%%%%%5
%
The Belle detector was a cylindrical large-solid-angle magnetic
spectrometer located at the interaction point of the KEKB  asymmetric energy $e^{+}e^{-}$ collider \cite{kekb}.
% (8.0 GeV $e^{-}$ and 3.5 GeV $e^{+}$). 
The detector consisted of a silicon vertex detector,
a central drift chamber, an array of aerogel threshold Cherenkov counters,  
a barrel-like arrangement of time-of-flight
scintillation counters, and an electromagnetic calorimeter (ECL)
composed of CsI(Tl) crystals located inside 
a super-conducting solenoid coil that provided a 1.5~T
axial magnetic field.  An iron flux-return located outside the coil was instrumented to detect $K_L^0$ mesons and to identify
muons. A detailed description of the detector can be found in Ref.~\cite{Belle}.

% The \mbox{Belle II} experiment is located at the SuperKEKB $e^{+}e^{-}$ collider~\cite{Super}. 
% The energies of electron and positron beams are 7.0 GeV and 4.0 GeV, respectively. 
The \mbox{Belle II} detector~\cite{detector}, located at the SuperKEKB $e^{+}e^{-}$ collider~\cite{Super}, is an upgraded version of the Belle detector. Belle II includes a silicon vertex
detector consisting of pixel sensors and double-sided strip detectors, and a central drift chamber. The central drift chamber is surrounded by two types of Cherenkov light detector systems: time-of-propagation detectors for the barrel region and an aerogel ring-imaging Cherenkov detector for the forward end cap region. The Belle ECL is reused in \mbox{Belle II} along with the solenoid and the iron flux return yoke. However, the ECL readout
electronics has been upgraded \cite{detector}. The solenoid flux return is instrumented with resistive-plate chambers and plastic scintillator modules to detect muons, $K^{0}_{L}$ mesons, and neutrons.

The $z$ axis of the laboratory frame is defined as the solenoid axis, where the positive direction is approximately that of the electron beam. This convention applies both to Belle and \mbox{Belle II}.

Monte Carlo simulated events are used to optimize the selection criteria, estimate signal selection efficiencies, train multivariate discriminants, identify various sources of background, and develop a model to fit data. We examine the data after all the requirements are fixed to avoid experimenter's bias. To study the signal, we use $10^{5}$ $\rm \Upsilon(4S) \to$ $B^{0}\overline{B}^{0}$ simulated decays in which one $B^{0}$ meson decays as $B^{0} \to \gamma\gamma$ and the other decays according to known decay modes tabulated by the Particle Data Group \cite{pdg}. The simulated signal and $e^{+}e^{−} \to B\overline{B}$ samples are generated using the \resizebox{!}{0.7em}{E}\resizebox{!}{0.6em}{VT}\resizebox{!}{0.7em}{G}\resizebox{!}{0.6em}{EN} \cite{evt} and \resizebox{!}{0.7em}{P}\resizebox{!}{0.6em}{YTHIA} 8.2 \cite{pythia} software packages, and the detector response is simulated using    \resizebox{!}{0.6em}{GEANT}\resizebox{!}{0.65em}{3} \cite{geant} and \resizebox{!}{0.6em}{GEANT}\resizebox{!}{0.65em}{4} \cite{geant4} for Belle and \mbox{Belle II}, respectively.  Continuum $e^{+}e^{−} \to q\overline{q}$ background processes, where $q = u, d, s, c$ are generated by \resizebox{!}{0.7em}{P}\resizebox{!}{0.6em}{YTHIA} 6.4 for Belle \cite{pythia6.4}. In the case of \mbox{Belle II}, the KKMC \cite{kkmc} generator is used for hard scattering, followed by \resizebox{!}{0.7em}{P}\resizebox{!}{0.6em}{YTHIA} 8.2 \cite{pythia} for the hadronization process. In addition, a sample of $e^{+}e^{-}\to \tau^{+}\tau^{-}$ events is generated with the \resizebox{!}{0.57em}{TAUOLA} package \cite{taula}. The simulated samples of the aforementioned background events corresponding to 1 ab$^{−1}$ or more are used.
 % \textcolor{red}{We use a MC sample approximately six and three times larger than the data sample to study backgrounds in Belle and Belle II, respectively.}
 % Also, the large sample of rare MC data,  comprising both charged and mixed rare background events is used to investigate the contamination from other $B$ decays. 
 To validate the simulations of continuum processes, Belle (\mbox{Belle II}) has collected 89.5 (42.3) fb$^{-1}$ of data about 60 MeV below the $\rm \Upsilon(4S)$ peak. Experimental and simulated Belle data are converted into Belle II format \cite{b2bii} and processed with the Belle II software \cite{basf2,b2}. 
  % \mbox{Belle II} data and MC are processed with the \mbox{Belle II} analysis software framework (BASF2) \cite{basf2,b2}, whereas the Belle data and MC are converted into BASF2 format using the B2BII software package \cite{b2bii}. 
 % Therefore, the analysis software is the same for both data samples, but the selection criteria are different.

Candidate \mbox{$B^{0} \to \gamma\gamma$} decays are characterized by two nearly back-to-back highly energetic photons in the $e^{+}e^{-}$ center-of-mass (c.m.) frame, as $B^{0}$ mesons are produced
almost at rest. Photons are selected from isolated energy deposits (clusters) in the ECL that are not associated with charged particle trajectories (tracks). We select events containing at least two photons with energies in the range $1.4<E^{\ast}({\gamma})<3.4$ GeV,  where the asterisk denotes an observable in the $e^{+}e^{-}$ c.m.\,\,frame. Only ECL clusters with polar angle $\theta$ in the $33^{\circ} < \theta < 132^{\circ}$ barrel region are considered.
To reject background from merged photon clusters and neutral hadrons, we require that
$E_{9}/E_{25}(E_{9}/E_{21})  > 0.95$ for Belle (\mbox{Belle II}).
% ${\tt E_{9}/E_{25}}$ is the ratio of the energy deposited in an innermost $(3 \times 3)$ array of crystals compared to that deposited in the $(5 \times 5)$ array centred around the most energetic crystal.
In Belle, $E_{9}/E_{25}$ corresponds to the ratio of energy deposits between ($3\times 3$) and ($5\times 5$) crystals centered on the crystal with maximum energy, whereas in Belle II, $E_{21}$ corresponds to the energy of the ($5\times 5$) crystals, excluding the corners. To suppress clusters originating from neutral hadrons, the number of crystals with energy above 20 MeV in the clusters must exceed 15. To distinguish between photon and $K^{0}_{L}$ showers, we utilize a boosted decision tree (BDT) \cite{fast} trained using Zernike moments \cite{zer} as inputs. The classifier output must have a value above 0.75, which is 90\% efficient in
selecting the signal while rejecting 74\% of the background events. To reject out-of-time QED processes such as Bhabha scattering or $e^{+}e^{-}\to\gamma\gamma$, the ECL cluster hit time is required to be within a 2 
$\text{\textmu}$s window
around the beam crossing time. For Belle II, the photon signal time, calculated from the fitted time of the highest energy crystal's recorded waveform within the cluster, should not differ from the beam crossing time by more than 200 ns.

The $B^0 \!\rightarrow\! \gamma\gamma$
 signal candidates are reconstructed by combining two photon candidates and selected using the beam-constrained mass $M_{\rm bc}$ and energy difference $\Delta E$ defined as\\
\begin{equation}
 M_{\rm bc} = \sqrt{(E_{\rm beam}^{\ast}/c^{2})^2 - (p_{B^0}^{\ast}/c)^2},   
\end{equation}
\begin{equation}
  \Delta E = E_{B^0}^{\ast} - E_{\rm beam}^{\ast},  \end{equation}
where $E_{\rm beam}^{\ast}$ is the beam energy, $E_{B^0}^{\ast}$ and $p_{B^0}^{\ast}$ are the energy and momentum of the $B^0$ candidate, all calculated in the $e^+e^-$ c.m. frame. The signal events peak at the $B^{0}$ meson mass in the $M_{\rm bc}$ distribution and concentrate near zero in the $\Delta E$ distribution. Hence, the $B^{0}$ meson candidates are required to be in the range 5.24 $< M_{\rm bc} <$ 5.29 $\text{GeV}/c^{2}$ and −0.6 $< \Delta E <$ 0.2 GeV. The $\Delta E$ window is not centered around zero because of energy leakage from the ECL. No events with multiple $B^{0}$ candidates are found in the experimental or simulated signal data.
To reduce $e^{+}e^{-}\to q\overline{q}$ and $\tau^{+}\tau^{-}$ events, we require at least three tracks in the event and the ratio of the second and zeroth Fox Wolfram moments \cite{SFW} to be less than 0.7. This ratio is computed using the momenta of all charged and neutral particles within the event.  
% To remove low multiplicity and continuum background events, we require at least two tracks in the event and the ratio of the second and zeroth Fox Wolfram moments \cite{SFW} to be smaller than 0.7. This ratio is computed using the momenta of all charged and neutral particles within the event. 
% Only photons in time with the rest of the event are retained. For Belle II, the timing criteria require the photon timing to be within 200 ns of the event time, and the absolute value of the ratio between {\tt clusterTiming} and {\tt clusterErrorTiming} should be less than 2.0. In this context, {\tt clusterTiming} represents the difference between the photon time and the event time, while {\tt clusterErrorTiming} indicates the error associated with the {\tt clusterTiming} variable.\\

% The significant background arises from photons produced by the asymmetric decays of $\pi^{0}$ or $\eta$ mesons into final states consisting of two gamma particles.
% Each high-energy photon candidate is paired with a low-energy photon from the event to create a $\pi^{0}$($\eta$) meson.
 
 To suppress the background from asymmetric-energy decays of $\pi^{0}\to\gamma\gamma$ and $\eta\to\gamma\gamma$ decays, a $\pi^{0}/\eta$ veto is implemented. The initial step is to pair each high-energy photon candidate with low-energy photon candidates having energies above 50 MeV in the event. The probability of a correctly reconstructed $\pi^{0}/\eta$ is then obtained using a BDT ($\rm BDT_{veto}$) trained on a dedicated sample dominated by $\pi^{0}$ and $\eta$ events. For training, we use a set of variables characterizing the photon pairs, including their invariant mass,  the cosine of the angle in
the $\pi^{0}/\eta$ rest frame between the momentum of the high-energy photon and the boost direction of $\pi^{0}/\eta$ from the laboratory frame. Other variables for the low-energy photon include its energy, its polar angle, the total number of crystal hits in the ECL, and the cluster variable $E_{9}/E_{21(25)}$. For Belle II, we use two additional variables: the distance between the ECL cluster and the nearest charged particle trajectory extrapolated to the ECL, and the output of a multivariate classifier based on Zernike moments for the low-energy photon. All the selection criteria described above are optimized by maximizing $\varepsilon_{\rm sig}/(3/2+\sqrt{\rm B})$ \cite{Punzi:638215}, where $\varepsilon_{\rm sig}$ is the efficiency of the event selection calculated from $B^{0}\to\gamma\gamma$ simulation, and B is the expected number of background events in the signal region. The optimized $\rm BDT_{veto}$ selection in Belle rejects 78\% of background photons from $\pi^{0}/\eta$ decays while retaining 85\% of the signal. Similarly, the \mbox{Belle II} selection rejects 75\% of this background while retaining 82\% of the signal.

The dominant source of the remaining background is from the continuum. Since light quarks carry significant momenta, continuum events are jetlike and are therefore topologically different from isotropic $B^{0}\overline{B}^{0}$ events in which $B^{0}$ meson pairs are produced nearly at rest in the c.m. frame. These differences in event shape topology provide most of the discrimination against continuum background. We separately train a $\rm BDT_{\it q\bar{q}}$ classifier using 21 variables for both Belle and \mbox{Belle II} data. The classification is based on modified Fox-Wolfram moments \cite{SFW}, the cosine of the angle between the thrust axis of the $B^{0}$
meson and the $z$ axis, the cosine of the angle between the thrust axis of the signal $B^{0}$ candidate and the
thrust axis of the rest of the event (the other $B^{0}$ in a correctly reconstructed event) \cite{costbto}, the cosine of the angle between the $B^{0}$ flight direction and the $z$-axis, the number of tracks in the event, and the flavor tagger output \cite{flavor}, which determines the flavor of the tag (non-signal) side $B^{0}$ meson. If the flavor tagger fails to find a tag for a $B^{0}$ candidate, which is generally the case for continuum background, the absence of $B^{0}$-flavor information is noted and input to $\rm BDT_{\it q\bar{q}}$. 

The BDT output, $C_{\text{BDT}_{q\bar{q}}}$, ranges from 0 to 1, with a value near 1(0) being more likely for a signal (background) event. We use simulated samples to determine a minimum threshold on the continuum classifier output that minimizes the average expected statistical uncertainty of the signal yield.
% The choice of the selection criterion on $C_{\text{BDT}}$ is determined based on Punzi's figure-of-merit (FOM) \cite{Punzi:638215}. 
This selection requires $C_{\text{BDT}_{q\bar{q}}} > 0.55$ $(C_{\text{BDT}_{q\bar{q}}} > 0.45)$, which rejects 93\% (87\%) of the $q\overline{q}$ background and retains 86\% (89\%) of the signal when applied to the Belle (\mbox{Belle II}) simulation. After applying all the selection criteria, the overall signal reconstruction efficiency for Belle and \mbox{Belle II} is ($23.3\pm0.1$)\% and ($30.8\pm0.1$)\%, respectively, where the uncertainties are statistical.

In addition, we evaluate the impact of $B^{0}\to\pi^{0}\pi^{0}$, $B^{0}\to\eta\eta$, $B^{0}\to\eta\pi^{0}$, and $B^{0}\to\omega\gamma$ background events. The largest contribution, from $B^{0}\to\pi^{0}\pi^{0}$ decay, constitutes 0.03 events. Therefore, we conclude these rare $B$ decay backgrounds are negligible.

To extract the signal yield, we perform a three-dimensional extended unbinned maximum likelihood fit to $M_{\text{bc}}$, $\Delta E$, and $C'_{\text{BDT}}$, where $C'_{\text{BDT}}$ is the output of the BDT ($C_{\text{BDT}_{q\bar{q}}}$) transformed using the probability integral transformation \cite{integral}. The $C'_{\text{BDT}}$ distribution for the simulated signal is uniform between zero and one, whereas the background distribution exhibits a peak at zero, simplifying modeling. The likelihood function is defined as
% the $C_{\text{BDT}}$ transformed following "The Probability Integral Transform" \cite{integral}. The $C'_{\text{BDT}}$ distribution for the simulated signal is uniform between zero and one, whereas the background distribution exhibits a peak at zero, simplifying modeling. The likelihood function is defined as 
\vspace{-2mm}
\begin{equation}
 \mathcal{L}_{\rm fit}= e^{-\sum\limits_{j} n_{j}}\prod_{i}^{N}(\sum\limits_{j} n_{j}P_{j}((M_{\text{bc}})^{i},(\Delta E)^{i},(C'_{\text{BDT}})^{i} )),
 \label{lfit}
\end{equation}
where $P_{j}$\,$(M_{\text{bc}},\Delta E, C'_{\text{BDT}})$ is the probability density function (PDF) of the signal or background component (specified by index $j$), $n_{j}$ is the yield of this component, $i$
represents the event index, and $N$ is the total number of events in the sample. The PDFs for the signal and background are based on simulation.
To model the $M_{\text{bc}}$ and $\Delta E$ signal distribution, a two-dimensional kernel-density \cite{ker} shape is employed to account for the correlation between $M_{\text{bc}}$ and $\Delta E$, which is 26\% and 17\% for Belle and Belle II, respectively. The distribution of $C'_{\text{BDT}}$ for the signal is modeled by a constant function. The background is characterized by an ARGUS function \cite{argus} for the $M_{\text{bc}}$ distribution, while the $\Delta E$ distribution is modeled using a first-order Chebychev polynomial. The $C'_{\text{BDT}}$ distribution is modeled by a sum of two exponential functions. All the signal parameters, $C'_{\text{BDT}}$ background parameters, and the ARGUS endpoint are fixed to the best-fit values obtained from one-dimensional fits to simulated events. All other background shape parameters and the signal and background yields are allowed to vary in the fit.

\vspace{2mm}
To test for bias in the fitted signal yield, we perform ensemble tests for Belle and Belle II for different signal yields. Signal events are randomly selected from simulation, while the expected number of background events are generated from the nominal PDFs. Statistical fluctuations in the number of signal and background events are included, assuming that the events follow Poisson statistics. Each simulated experiment is repeated 1000 times for signal yield ranging from 2 to 20 events and fitted with the nominal model as defined in Eq.~(\ref{lfit}). The average deviation between the fitted and generated signal yields is treated as a source of systematic uncertainty. Additionally, we include the linearity of fit results relative to signal yield as an additional source of systematic uncertainty. Combined in quadrature, this yields a +0.14 (+0.10) uncertainty on the fit bias for Belle (\mbox{Belle II}).

\vspace{2mm}
The systematic uncertainty associated with fixing the parameter values of the PDFs is assessed by varying the best-fit parameter values within $\pm1\sigma$ of their statistical uncertainties. The resulting deviations in the signal yields in data are measured and used to quantify this uncertainty. A systematic uncertainty of $^{+0.56}_{-0.48}$ ($^{+0.28}_{-0.32}$) events is assigned to the fit model for Belle (\mbox{Belle II}).
To evaluate the accuracy of the simulation in describing data distributions, we compare its predictions to data from a $B^{0}\to K^{\ast}(892)^{0}\gamma$ control sample. The $K^{\ast}(892)^{0}$ mesons are reconstructed using $K^{\ast}(892)^{0} \to K^{+}\pi^{−}$ decays, in which the charged kaon is required to have $R_{K/\pi}= \mathcal{L}_{K} /(\mathcal{L}_{K} + \mathcal{L}_{\pi} ) >$ 0.6, where $\mathcal{L}_{K(\pi)}$ is the likelihood for the kaon(pion) hypothesis, which combines information from various subdetectors of Belle or \mbox{Belle II}. The photon selection 
criteria are the same as in the signal reconstruction. In addition, the invariant mass of the $K^{+}\pi^{-}$ meson pair should be in the range $0.817 < M_{K^{\ast}} < 0.968 $ GeV/$c^{2}$. The deviations from unity in the data/MC ratio for the distributions of $M_{\rm bc}$ and $\Delta E$ are considered as a source of systematic uncertainty. By adding them in quadrature, we assign a systematic uncertainty of +0.06 (+0.04) events for Belle (Belle II) as the signal shape un-

% decays}. For Belle (Belle II), we find a deviation of +0.06 (+0.04) events due to the signal shape in the control sample.}
% These uncertainties are combined in quadrature,
\newpage
\onecolumngrid

\begin{figure}
	\centering
 \includegraphics[width=0.32\linewidth]{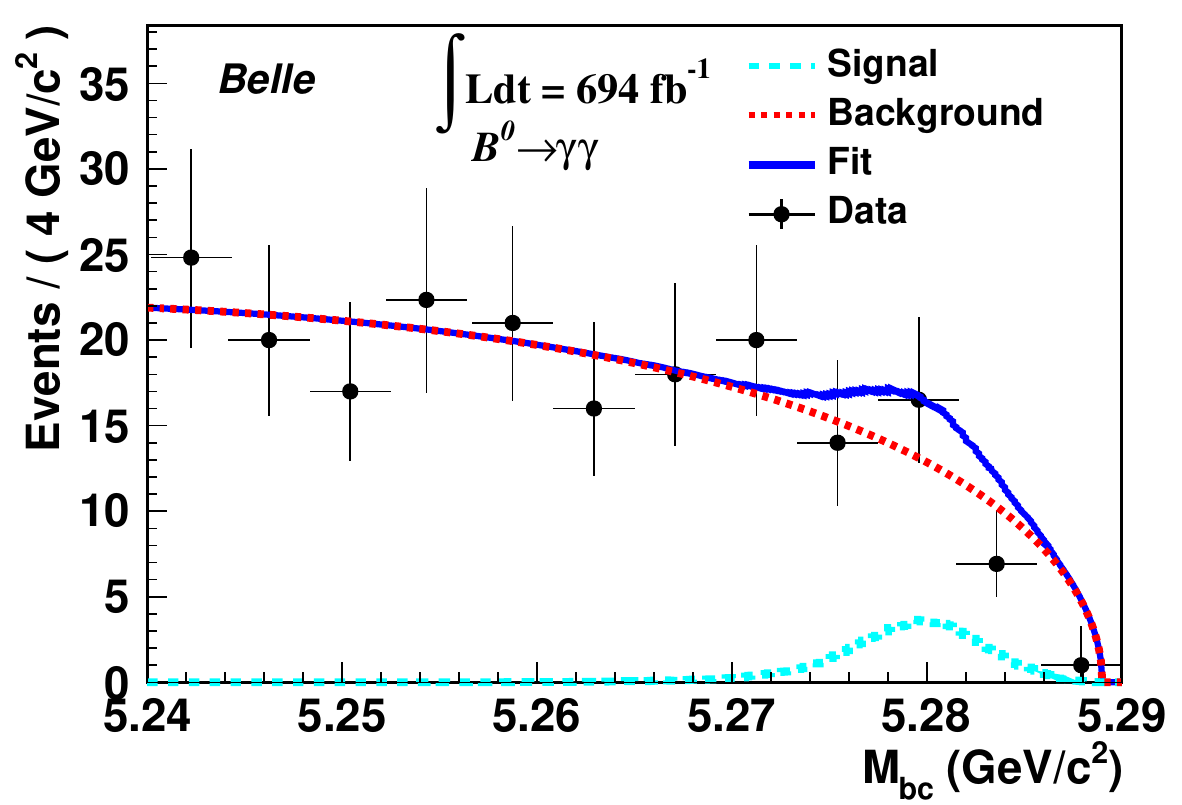}
 \includegraphics[width=0.32\linewidth]{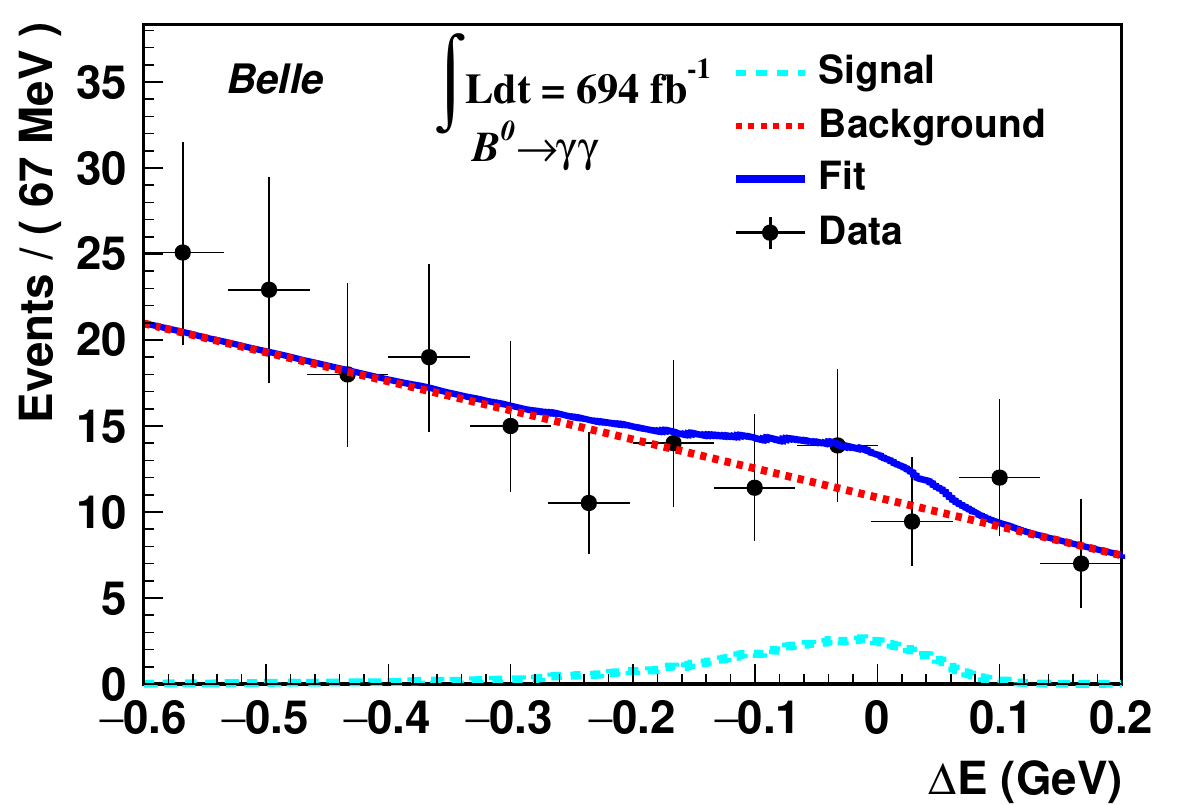}
  \includegraphics[width=0.32\linewidth]{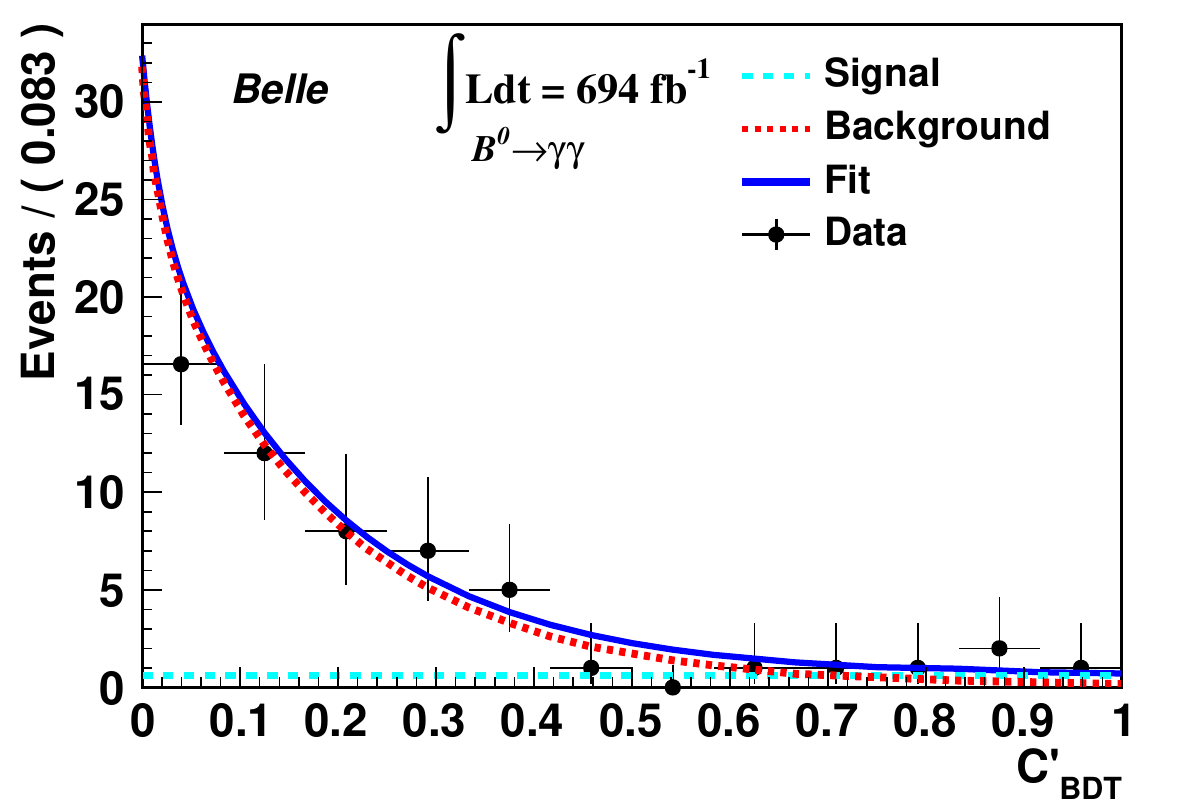}
  \includegraphics[width=0.32\linewidth]{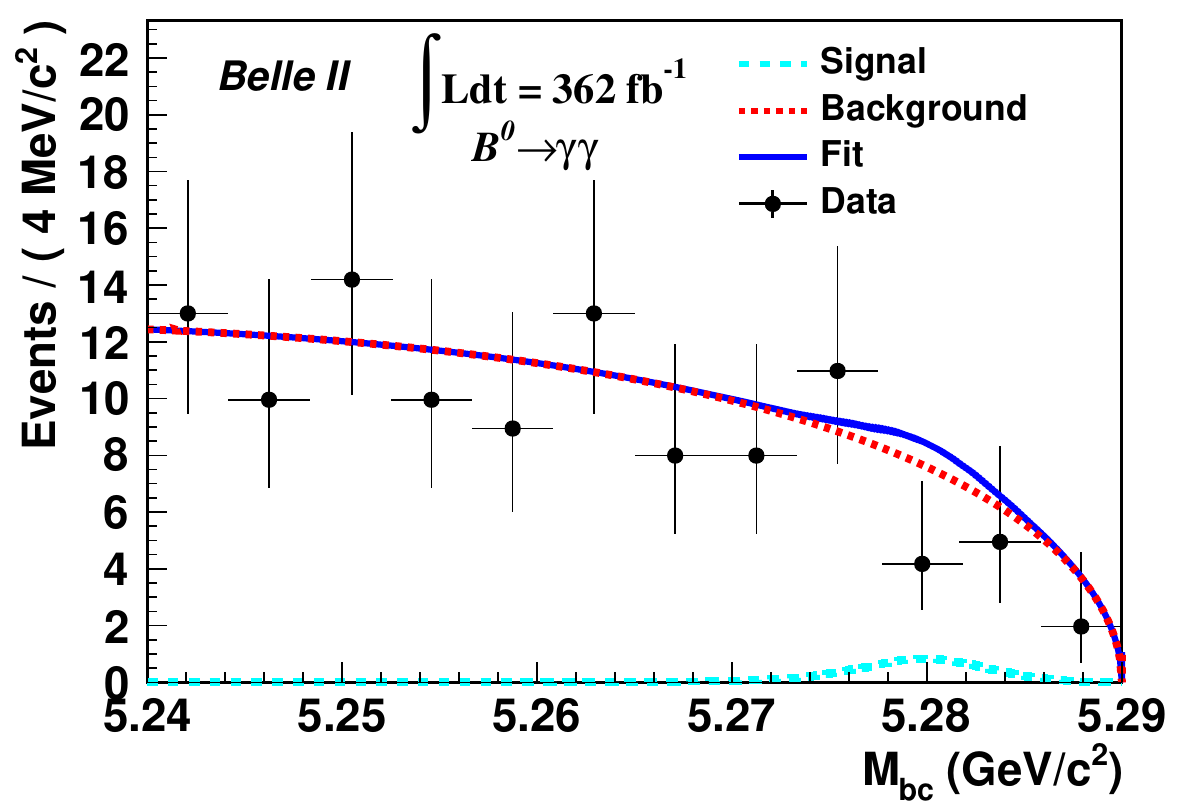}
 \includegraphics[width=0.32\linewidth]{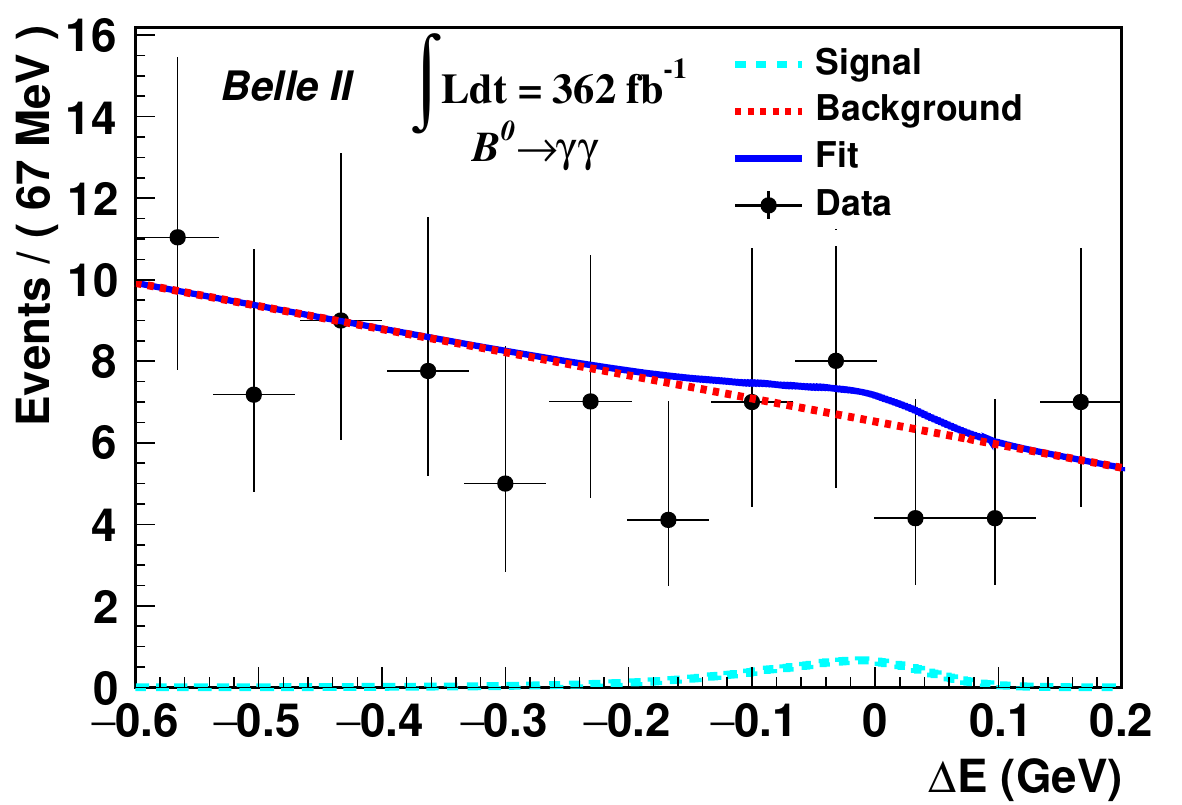}
  \includegraphics[width=0.32\linewidth]{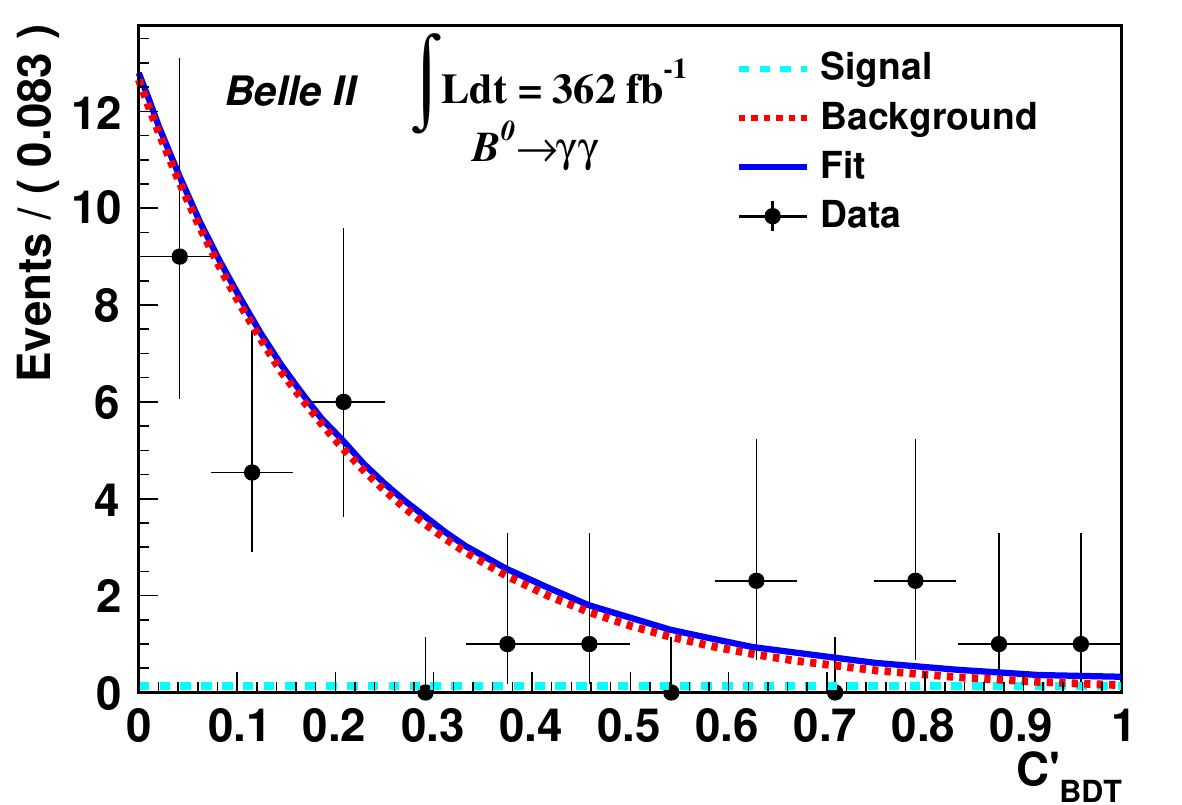}
\caption{Signal enhanced projections of $M_{\text{bc}}$ (left), $\Delta E$ (middle), and $C'_{\text{BDT}}$ (right) for the $B^{0}\to \gamma \gamma$ analysis using the Belle (top) and \mbox{Belle II} (bottom) dataset. For each plot, we apply the signal region selection criteria on the variables other than the plotted variable. The signal regions for the first two variables are as follows, 5.27 $<$  $M_{\text{bc}}$ $<$ 5.29  $\text{GeV}/c^{2}$ and −0.19 $<$ $\Delta E$ $<$ 0.14 GeV for Belle and 5.27 $<$  $M_{\text{bc}}$ $<$ 5.29  $\text{GeV}/c^{2}$ and −0.19 GeV $<$ $\Delta E$ $<$ 0.15 GeV for \mbox{Belle II}. The cyan(dashed), red(dotted), and blue(solid) color distributions represent the signal, continuum background, and total fit function, respectively. Points with error bars represent data. }
% The difference between the observed and fit values divided by the uncertainty from the fit (pulls) are shown below each distribution.} 
\label{bellefit}
\end{figure}
\twocolumngrid

\begin{table}[htb]
\caption{Summary of additive systematic uncertainties.}
\label{sys1}
% \begin{tabular}{p{4.5cm}p{1.5 cm}p{1.2cm}}
\begin{tabular}{p{4.4cm}p{1.1 cm}p{1.1cm}p{1.1 cm}}
\hline \hline
% \multicolumn{3}{c}{} \\
% \hline
Source & Belle (events) & \mbox{Belle II} (events) & Combined (events)\\
\hline
Fit bias & $+0.14$ &$+0.10$ & +0.12\\
PDF parametrization & $^{+0.56}_{-0.48}$ & $^{+0.28}_{-0.32}$ & $^{+0.52}_{-0.44}$\\
Shape modeling & +0.06 & +0.04 & +0.05 \\
\hline
Total (sum in quadrature) & $^{+0.58}_{-0.48}$ & $^{+0.30}_{-0.32}$ & $^{+0.54}_{-0.44}$\\
\hline \hline
\end{tabular}
\end{table}
\vspace{-0.1cm}
\begin{table}[htb]
\caption{Summary of multiplicative systematic uncertainties.}

\begin{tabular}{p{4.5cm}p{1.1cm}p{1.1cm}p{1.1cm}}
\hline \hline
% \multicolumn{3}{c}{Multiplicative systematic uncertainties} \\
% \hline
Source & Belle (\%) & \mbox{Belle II} (\%) & Combined (\%)\\
\hline
Photon detection efficiency &$4.0$ & 2.7 & 3.5\\
Simulation sample size &0.4 &0.3 & 0.3\\
Number of $B\bar{B}$ & 1.3& 1.5 & 1.0\\
$f^{00}$ & 2.5 & 2.5 & 2.5\\
$C_{\text{BDT}}$ requirement& 0.4 & 0.9 & 0.6\\
$\pi^{0}/\eta$ veto & $0.4$ & 0.6 & 0.4\\
Timing requirement efficiency & 2.8 & \ $-$ & 2.7\\
\hline
Total (sum in quadrature) & 5.7 & 4.1 & 5.2\\
\hline \hline
\end{tabular}
\label{sys2}
\end{table}

\noindent certainty. These uncertainties are combined in quadrature, resulting in a systematic 
uncertainty of \mbox{$^{+0.58}_{-0.48}$} ($^{+0.30}_{-0.32}$) events, as presented in Table \ref{sys1}. These uncertainties are treated as additive systematic uncertainties that affect the significance of the observed signal yield. Table \ref{sys2} includes uncertainties in the photon detection efficiency, the signal reconstruction efficiency, the number of produced $B\overline{B}$ pairs, and the
branching fraction of $\rm \Upsilon(4S)$ to neutral $B\overline{B}$ pairs, $f^{00}$ \cite{f00}. These uncertainties are multiplicative, which are proportional to the signal yield and affect the signal efficiency in the denominator of Eq. (\ref{eq2}).
The systematic uncertainty arising from the photon detection efficiency is determined to be 4.0\% for Belle using the recoil technique in $e^{+}e^{−} \to e^{+}e^{−}\gamma$ radiative Bhabha events. For Belle II data, it is measured to be 2.7\% utilizing a $e^{+}e^{−} \to \mu^{+}\mu^{-}\gamma$ initial-state radiation data sample. The uncertainty in signal reconstruction efficiency is due to the limited size of the signal simulated sample, and is determined to be 0.4\% (0.3\%) for Belle (\mbox{Belle II}). The uncertainties on the number of $B\overline{B}$ pairs recorded in Belle and \mbox{Belle II} are also considered.
% The systematic uncertainty in photon detection efficiency is determined as 4.0\% for Belle using the recoil technique in radiative Bhabha events $e^{+}e^{−} \to e^{+}e^{−}\gamma$, while Belle II measures it to be 2.7\% utilizing a $e^{+}e^{−} \to \mu^{+}\mu^{-}\gamma$ data sample, which accounts for initial-state radiation. The uncertainty in signal reconstruction efficiency, attributed to MC sample size, is determined to be 0.6\% (0.5\%) for Belle (Belle II). % The uncertainties on the number of $B\bar{B}$ pairs recorded in Belle and Belle II are also considered. The uncertainties on the $f^{00}$ are also included.
 A systematic uncertainty from the efficiency of the requirement on $C_{\text{BDT}}$ and the $\pi^{0}/\eta$ veto is estimated using the $B^{0}\to K^{\ast}(892)^{0}\gamma$ control sample. The efficiency ratio between the data and simulation of those requirements is used as correction, and its uncertainty as the associated systematic error. Since there are two photons in the final state for signal candidates and only one in the control mode, the correction and systematic uncertainty are doubled.
  % Additionally, in the $C_{\text{BDT}}$, a $K^{\ast}(890)^{0}$ is treated as a photon.
  The efficiency ratio between the data and simulation due to the $C_{\text{BDT}}$ and the $\pi^{0}/\eta$ veto requirements for Belle (Belle II) are 1.026$\pm$0.004 (1.006$\pm$0.009) and 1.004$\pm$0.004 (1.002$\pm$0.006), respectively.
An uncertainty of 2.8\% is assigned due to the timing criteria for Belle, while for \mbox{Belle II}, the uncertainty is incorporated into the photon detection efficiency.
% \begin{table}[htb]
% \caption{Summary of additive systematic uncertainties.}
% \label{sys1}
% % \begin{tabular}{p{4.5cm}p{1.5 cm}p{1.2cm}}
% \begin{tabular}{p{4.4cm}p{1.1 cm}p{1.1cm}p{1.1 cm}}
% \hline \hline
% % \multicolumn{3}{c}{} \\
% % \hline
% Source & Belle (events) & \mbox{Belle II} (events) & Combined (events)\\
% \hline
% Fit bias & $+0.14$ &$+0.10$ & +0.12\\
% PDF parameterization & $^{+0.56}_{-0.48}$ & $^{+0.28}_{-0.32}$ & $^{+0.52}_{-0.44}$\\
% Shape modeling & +0.06 & +0.04 & +0.05 \\
% \hline
% Total (sum in quadrature) & $^{+0.58}_{-0.48}$ & $^{+0.30}_{-0.32}$ & $^{+0.54}_{-0.44}$\\
% \hline \hline
% \end{tabular}
% \end{table}
%  \vspace{-0.1cm}
% \begin{table}[htb]
% \caption{Summary of multiplicative systematic uncertainties.}

% \begin{tabular}{p{4.5cm}p{1.1cm}p{1.1cm}p{1.1cm}}
% \hline \hline
% % \multicolumn{3}{c}{Multiplicative systematic uncertainties} \\
% % \hline
% Source & Belle (\%) & \mbox{Belle II} (\%) & Combined (\%)\\
% \hline
% Photon detection efficiency &$4.0$ & 2.7 & 3.5\\
% MC statistics &0.4 &0.3 & 0.3\\
% Number of $B\bar{B}$ & 1.3& 1.5 & 1.0\\
% $f^{00}$ & 2.5 & 2.5 & 2.5\\
% $C_{\text{BDT}}$ requirement& 0.4 & 0.9 & 0.6\\
% $\pi^{0}/\eta$ veto & $0.4$ & 0.6 & 0.4\\
% Timing requirement efficiency & 2.8 & \ $-$ & 2.7\\
% \hline
% Total (sum in quadrature) & 5.7 & 4.1 & 5.2\\
% \hline \hline
% \end{tabular}
% \label{sys2}
% \end{table}

We obtain 9.1$^{+5.6}_{-4.4}$ (1.9$^{+4.2}_{-2.8}$) signal events and 615$\pm$25 (317$\pm$18) background events for Belle (\mbox{Belle II}) from the fits to the two independent datasets.
The branching fraction is calculated using the equation
\begin{equation}
  \mathcal{B}(B^{0} \to\gamma\gamma) = \frac{N^{\rm fit}_{\rm sig}}{ 2 \times N_{B\overline{B}}\times   \epsilon_{\rm rec}\times f^{00}}, 
  \label{eq2}
\end{equation}
where $N^{\rm fit}_{\rm sig}$ represents the signal yield obtained from the fit, $N_{B\overline{B}} = (753 \pm 10) \times 10^{6}$ and $(387 \pm 6) \times 10^{6}$ is the number of $B\overline{B}$ pairs at the $\rm \Upsilon(4S)$ resonance for Belle and \mbox{Belle II}, $\epsilon_{\rm rec}$ = 23.3\% and 30.8\% are the signal reconstruction efficiencies for Belle and \mbox{Belle II}, respectively, and $f^{00}= (48.4 \pm 1.2)\%$. Therefore, the branching fractions for the Belle and \mbox{Belle II} datasets are $(5.4^{+3.3}_{-2.6} \pm 0.5)\times 10^{-8}$ and $(1.7^{+3.7}_{-2.4} \pm 0.3)\times 10^{-8}$, respectively. The first uncertainty is statistical, while the second is systematic.

In addition, we perform an extended unbinned maximum likelihood fit to the $M_{\text{bc}}$, $\Delta E$, and $C'_{\text{BDT}}$ distributions simultaneously in the Belle and Belle II datasets, which is shown in Fig. \ref{bellefit}. The branching fraction is determined to be $(3.7^{+2.2}_{-1.8}\pm 0.5)\times 10^{-8}$ with a total signal (background) yield of 11.0$^{+6.5}_{-5.5}$ (931$\pm$31) events, where the uncertainties are statistical only. The combined systematic uncertainty is calculated as the maximum deviation between the fitted values and the best-fit values with the inclusion of systematic uncertainty in the simultaneous fit. The signal significance is calculated as $\sqrt{-2\ln(\mathcal{L}_{0}/\mathcal{L}_{\rm max})}$, where $\mathcal{L}_{0}$ is the maximum value of the likelihood when signal yield is fixed to zero, and $\mathcal{L}_{\rm max}$ is the maximum value of the likelihood of the nominal fit. The resulting significance is 2.5$\sigma$, which includes the systematic uncertainties. To include systematic uncertainties in the significance, we convolve the likelihood distribution
with a Gaussian function whose width is set to the total systematic uncertainty.

As the significance of the signal yield is low, we calculate an upper limit (UL) on the $\mathcal{B}$ using a Bayesian approach, with a flat prior.
The UL on the branching fraction is determined by integrating the likelihood function including the systematic uncertainty from zero to 90\% of the area under the curve. 
% The modified ratio is then re-convolved with a Gaussian function of width proportional to the signals, where the total multiplicative systematic uncertainty is the proportionality constant. 
The upper limit on the branching fraction obtained from the combined dataset is $6.4\times 10^{-8}$, at 90\% credibility level \cite{cl}. The expected upper limit from the simulation is $4.4\times 10^{-8}$ at 90\% credibility level. The measured branching fractions and the resulting upper limits on $\mathcal{B}(B^{0} \to \gamma\gamma)$ at 90\% credibility level, including the systematic uncertainties, are summarized in Table \ref{ul}.

\begin{table}[htb]
\caption{Summary of $\mathcal{B}(B^{0}\to\gamma\gamma)$ measurements and UL's at 90\% credibility level.}

\begin{tabular}{p{1.7cm}p{3.6cm}p{2.8cm}}
\hline \hline
 
 & \ \ \ \ \ $\mathcal{B}(B^{0}\to\gamma\gamma)$ & 
    UL on $\mathcal{B}(B^{0}\to\gamma\gamma)$ 
    \\
    & & \\
\hline
Belle &$(5.4^{+3.3}_{-2.6}\pm0.5)\times10^{-8}$  & \ \ $< 9.9\times10^{-8}$ \\
Belle II & $(1.7^{+3.7}_{-2.4}\pm0.3)\times10^{-8}$  & \ \ $< 7.4\times10^{-8}$ \\
Combined &$(3.7^{+2.2}_{-1.8}\pm0.5)\times10^{-8}$ & \ \ $< 6.4\times10^{-8}$\\

\hline \hline
\end{tabular}
\label{ul}
\end{table}

% \vspace{0.4cm}
In summary, we have searched for the decay $B^{0} \to \gamma\gamma$ using a 1.1 ab$^{-1}$ data sample collected at the $\rm \Upsilon(4S)$ resonance by the Belle and \mbox{Belle II} experiments. No statistically significant signal is observed, leading us to set an upper limit of $6.4\times10^{-8}$ on the branching fraction at 90\% credibility level. 
This result supersedes the previous Belle measurement \cite{belleul} and represents a significant improvement over the previous searches by the \textit{B\resizebox{!}{0.6em}{A}B\resizebox{!}{0.6em}{AR}} and Belle collaborations. The use of advanced analysis techniques such as BDTs results in a factor of two background reduction compared to the \textit{B\resizebox{!}{0.6em}{A}B\resizebox{!}{0.6em}{AR}} results and a gain of a factor of two in the signal reconstruction efficiency compared to the previous Belle measurements. These improvements, combined with the larger Belle + Belle II dataset, lead to an UL that is five times more restrictive than the previous best limit from \textit{B\resizebox{!}{0.6em}{A}B\resizebox{!}{0.6em}{AR}} \cite{del2011search}.\\

\section{ACKNOWLEDGMENTS}
% Policy from October 20, 2022
This work, based on data collected using the Belle II detector, which was built and commissioned prior to March 2019,
and data collected using the Belle detector, which was operated until June 2010,
was supported by
%Armenia
Higher Education and Science Committee of the Republic of Armenia Grant No.~23LCG-1C011;
%Australia
Australian Research Council and Research Grants
No.~DP200101792, % Jackson
No.~DP210101900, % Urquijo
No.~DP210102831, % Sevior
No.~DE220100462, % Hsu
No.~LE210100098, % Infrastructure
and
No.~LE230100085; % Infrastructure
%Austria
Austrian Federal Ministry of Education, Science and Research,
Austrian Science Fund
No.~P~34529,
No.~J~4731,
No.~J~4625,
and
No.~M~3153,
and
Horizon 2020 ERC Starting Grant No.~947006 ``InterLeptons'';
%Canada
Natural Sciences and Engineering Research Council of Canada, Compute Canada and CANARIE;
%China
National Key R\&D Program of China under Contract No.~2022YFA1601903,
National Natural Science Foundation of China and Research Grants
No.~11575017,
No.~11761141009,
No.~11705209,
No.~11975076,
No.~12135005,
No.~12150004,
No.~12161141008,
and
No.~12175041,
and Shandong Provincial Natural Science Foundation Project~ZR2022JQ02;
%Czech Republic
the Czech Science Foundation Grant No.~22-18469S 
and
Charles University Grant Agency project No.~246122;
%EU
European Research Council, Seventh Framework PIEF-GA-2013-622527,
Horizon 2020 ERC-Advanced Grants No.~267104 and No.~884719,
Horizon 2020 ERC-Consolidator Grant No.~819127,
Horizon 2020 Marie Sklodowska-Curie Grant Agreement No.~700525 ``NIOBE''
and
No.~101026516,
and
Horizon 2020 Marie Sklodowska-Curie RISE project JENNIFER2 Grant Agreement No.~822070 (European grants);
%France
L'Institut National de Physique Nucl\'{e}aire et de Physique des Particules (IN2P3) du CNRS
and
L'Agence Nationale de la Recherche (ANR) under grant ANR-21-CE31-0009 (France);
%Germany
BMBF, DFG, HGF, MPG, and AvH Foundation (Germany);
%India
Department of Atomic Energy under Project Identification No.~RTI 4002,
Department of Science and Technology,
and
UPES SEED funding programs
No.~UPES/R\&D-SEED-INFRA/17052023/01 and
No.~UPES/R\&D-SOE/20062022/06 (India);
%Israel
Israel Science Foundation Grant No.~2476/17,
U.S.-Israel Binational Science Foundation Grant No.~2016113, and
Israel Ministry of Science Grant No.~3-16543;
%Italy
Istituto Nazionale di Fisica Nucleare and the Research Grants BELLE2;
%Japan
Japan Society for the Promotion of Science, Grant-in-Aid for Scientific Research Grants
No.~16H03968,
No.~16H03993,
No.~16H06492,
No.~16K05323,
No.~17H01133,
No.~17H05405,
No.~18K03621,
No.~18H03710,
No.~18H05226,
No.~19H00682, % Niigata
No.~20H05850,
No.~20H05858,
No.~22H00144,
No.~22K14056,
No.~22K21347,
No.~23H05433,
No.~26220706,
and
No.~26400255,
%the National Institute of Informatics, and Science Information NETwork 5 (SINET5), 
and
the Ministry of Education, Culture, Sports, Science, and Technology (MEXT) of Japan;  
%Korea
National Research Foundation (NRF) of Korea Grants
No.~2016R1\-D1A1B\-02012900,
No.~2018R1\-A2B\-3003643,
No.~2018R1\-A6A1A\-06024970,
No.~2019R1\-I1A3A\-01058933,
No.~2021R1\-A6A1A\-03043957,
No.~2021R1\-F1A\-1060423,
No.~2021R1\-F1A\-1064008,
No.~2022R1\-A2C\-1003993,
and
No.~RS-2022-00197659,
Radiation Science Research Institute,
Foreign Large-Size Research Facility Application Supporting project,
the Global Science Experimental Data Hub Center of the Korea Institute of Science and Technology Information
and
KREONET/GLORIAD;
%Malaysia
Universiti Malaya RU grant, Akademi Sains Malaysia, and Ministry of Education Malaysia;
%Mexico
% CINVESTAV-IPN, UNAM, UAS, BUAP and CONACYT are funded under
Frontiers of Science Program Contracts
No.~FOINS-296,
No.~CB-221329,
No.~CB-236394,
No.~CB-254409,
and
No.~CB-180023, and SEP-CINVESTAV Research Grant No.~237 (Mexico);
%Poland
the Polish Ministry of Science and Higher Education and the National Science Center;
%Russia
the Ministry of Science and Higher Education of the Russian Federation
and
the HSE University Basic Research Program, Moscow;
%Saudi Arabia
University of Tabuk Research Grants
No.~S-0256-1438 and No.~S-0280-1439 (Saudi Arabia);
%Slovenia
Slovenian Research Agency and Research Grants
No.~J1-9124
and
No.~P1-0135;
%Spain
Ikerbasque, Basque Foundation for Science,
the State Agency for Research of the Spanish Ministry of Science and Innovation through Grant No. PID2022-136510NB-C33,
Agencia Estatal de Investigacion, Spain
Grant No.~RYC2020-029875-I
and
Generalitat Valenciana, Spain
Grant No.~CIDEGENT/2018/020;
%Swiss (Belle 1)
the Swiss National Science Foundation;
%Sweden
The Knut and Alice Wallenberg Foundation (Sweden), Contracts No.~2021.0174 and No.~2021.0299;
%Taiwan
National Science and Technology Council,
and
Ministry of Education (Taiwan);
%Thailand
Thailand Center of Excellence in Physics;
%Turkey
TUBITAK ULAKBIM (Turkey);
%Ukraine
National Research Foundation of Ukraine, Project No.~2020.02/0257,
and
Ministry of Education and Science of Ukraine;
%USA
the U.S. National Science Foundation and Research Grants
No.~PHY-1913789 % Indiana CEEM
and
No.~PHY-2111604, % Luther
and the U.S. Department of Energy and Research Awards
No.~DE-AC06-76RLO1830, % PNNL
No.~DE-SC0007983, % Wayne State
No.~DE-SC0009824, % Florida
No.~DE-SC0009973, % VPI
No.~DE-SC0010007, % Duke
No.~DE-SC0010073, % South Carolina
No.~DE-SC0010118, % Carnegie Mellon
No.~DE-SC0010504, % Hawaii
No.~DE-SC0011784, % Cincinnati
No.~DE-SC0012704, % BNL
No.~DE-SC0019230, % Duke
No.~DE-SC0021274, % Mississippi
No.~DE-SC0021616, % Mississippi
No.~DE-SC0022350, % Louisville
No.~DE-SC0023470; % South Alabama
%last group
and
%Vietnam
the Vietnam Academy of Science and Technology (VAST) under Grants
No.~NVCC.05.12/22-23
and
No.~DL0000.02/24-25.

% Policy from October 20, 2022
These acknowledgements are not to be interpreted as an endorsement of any statement made
by any of our institutes, funding agencies, governments, or their representatives.

We thank the SuperKEKB team for delivering high-luminosity collisions;
the KEK cryogenics group for the efficient operation of the detector solenoid magnet;
the KEK Computer Research Center for on-site computing support; the NII for Science Information NETwork 6 (SINET6) network support;
and the raw-data centers hosted by BNL, DESY, GridKa, IN2P3, INFN, 
PNNL/EMSL, 
and the University of Victoria.


\begin{thebibliography}{99}
\bibitem{cl}
For Bayesian upper limits, it is now standard to quote credibility intervals rather than confidence levels, which are appropriate for frequentist analysis \cite{credibilty}.
\end{thebibliography}
\end{document}